\documentclass[sigconf, nonacm]{acmart}

\setcopyright{none}

\usepackage[utf8]{inputenc}
\usepackage{textcomp}

\usepackage{amsfonts}

\usepackage{subcaption}
\usepackage{listings}

\usepackage[linesnumbered, ruled, vlined]{algorithm2e}

\usepackage{booktabs}  % publication quality tables

\usepackage[inline]{enumitem}  % enhanced enumerations
\usepackage{siunitx}  % correct units for physical quantities

\usepackage{hyperref}  % import as last package
\usepackage[capitalise]{cleveref}  % import after hyperref

\usepackage{multirow}

\newcommand{\affiliationqarnot}{\affiliation{%
	\institution{Qarnot Computing}
	\city{Montrouge}%\city{F-92120 Montrouge}
	\country{France}
}}
\newcommand{\affiliationcristal}{\affiliation{%
	\institution{Univ.\@ Lille, Inria, CNRS, Centrale Lille, UMR 9189 CRIStAL}
	\city{Lille}%\city{F-59000 Lille}
	\country{France}
}}
\newcommand{\affiliationlig}{\affiliation{%
	\institution{Univ.\@ Grenoble Alpes, Inria, CNRS, LIG}
	\city{Grenoble}%\city{F-38000 Grenoble}
	\country{France}
}}
\newcommand{\affiliationgipsa}{\affiliation{%
	\institution{Univ.\@ Grenoble Alpes, CNRS, Grenoble INP, GIPSA-lab}
	\city{Grenoble}%\city{F-38000 Grenoble}
	\country{France}
}}

\begin{document}

\title{Mitigating Shared Storage Congestion Using Control Theory}

\author{Thomas \textsc{Collignon}}
\orcid{0009-0007-3902-0796}
\affiliationqarnot
\affiliationcristal
\email{thomas.collignon@qarnot-computing.com}

\author{Kouds \textsc{Halitim}}
\orcid{0009-0006-1043-0495}
\affiliationlig
\email{kouds.halitim@inria.fr}

\author{Raphaël \textsc{Bleuse}}
\orcid{0000-0002-6728-2132}
\affiliationlig
\email{raphael.bleuse@inria.fr}

\author{Sophie \textsc{Cerf}}
\orcid{0000-0003-0122-0796}
\affiliationlig
\email{sophie.cerf@inria.fr}

\author{Bogdan \textsc{Robu}}
\orcid{0000-0001-7568-007X}
\affiliationgipsa
\email{bogdan.robu@gipsa-lab.grenoble-inp.fr}

\author{Éric \textsc{Rutten}}
\orcid{0000-0001-8696-8212}
\affiliationlig
\email{eric.rutten@inria.fr}

\author{Lionel \textsc{Seinturier}}
\orcid{0000-0003-0006-6088}
\affiliationcristal
\email{lionel.seinturier@inria.fr}

\author{Alexandre \textsc{van Kempen}}
\orcid{0000-0001-8321-6779}
\affiliationqarnot
\email{alexandre.vankempen@qarnot-computing.com}

% redefine command for a more concise list of authors' names in the page headers
\renewcommand{\shortauthors}{%
    T.\@ \textsc{Collignon},
    K.\@ \textsc{Halitim},
    R.\@ \textsc{Bleuse},
    S.\@ \textsc{Cerf},
    B.\@ \textsc{Robu},
    É.\@ \textsc{Rutten},
    L.\@ \textsc{Seinturier},
    and
    A.\@ \textsc{van Kempen}
}

\begin{abstract}
    Efficient data access in High-Performance Computing (HPC) systems is essential to the performance of intensive computing tasks. Traditional optimizations of the I/O stack aim to improve peak performance but are often workload specific and require deep expertise, making them difficult to generalize
     %%%ER%%%
    or re-use. In shared HPC environments, resource congestion can lead to unpredictable performance, causing slowdowns and timeouts. To address these challenges, we propose a self-adaptive approach based on Control Theory to dynamically regulate client-side I/O rates. Our approach leverages a small set of runtime system load metrics to reduce congestion and enhance performance stability. We implement a controller in a multi-node cluster and evaluate it on a real testbed under a representative workload. Experimental results demonstrate that our method effectively mitigates I/O congestion, reducing total runtime by up to 20\% and lowering tail latency, while maintaining stable performance.
    
\end{abstract}




\maketitle

\section{Introduction}
    High-Performance Computing (HPC) systems play an increasingly crucial role in scientific discovery, complex simulations, fast prototyping, and other computationally intensive tasks.
    Performance in these systems heavily depends on efficient data access, making fast and scalable parallel I/O a critical requirement for many HPC applications.
    To meet this demand, the HPC I/O stack has grown highly complex, spanning multiple layers---including hardware, filesystems, and middleware---each with thousands of tunable parameters (see~\cite{io_stack_survey} for a comprehensive survey).
    Traditional optimization approaches focus on maximizing peak I/O performance through exhaustive parameter tuning, sophisticated benchmarking tools~\cite{filebench}, and, more recently, machine learning techniques~\cite{autotune, autotune2}. 
    While such tuning can yield significant performance gains, it is usually workload-specific, requires deep expertise, and does not generalize easily across diverse HPC environments.
    
    In the context of shared HPC infrastructures (e.g., cloud-based HPC), multiple users and applications may compete for limited storage resources, leading to resource congestion that can significantly degrade system-wide performance~\cite{sched_hpc_congestion}. 
    Storage congestion results in highly unpredictable performance, causing random slowdowns and timeouts. In many cases, unpredictability can be more problematic than consistently lower but stable performance. 
    In this context, we argue that preventing performance degradation due to resource congestion can sometimes be more important than maximizing peak I/O speed for individual applications. Several studies follow this approach by trying to mitigate I/O interferences~\cite{calciom, YildizO2016Root}. However, this requires an \emph{a priori} knowledge about the I/O patterns~\cite{io_patterns_survey} of the applications, which 
    %might not always be 
    is hardly
    available.
    
    To mitigate congestion, we propose a coarse-grained, self-adaptive approach that relies on control theory. 
    Rather than fine-tuning each layer of the I/O stack, our method dynamically regulates client-side I/O rates using real-time system load metrics. 
    By formulating the problem within a feedback-control framework, we abstract away much of the stack’s inherent complexity. 
    This paper makes the following contributions:
    \begin{itemize}
        \item We present a generic control-theoretic architecture for designing a controller that employs an actuator--sensor pair to dynamically adjust load on shared storage, ensuring more stable and predictable performance.
        \item We describe the implementation of a
        simple %%%ER%%%
        proportional--integral (PI) feedback controller on a computing cluster, including guidelines for tuning it to achieve effective regulation and noise reduction.
        \item We provide experimental results from a deployment on a real testbed with a representative write-intensive workload. 
        \item We discuss 
        paths we have identified 
        %%%ER%%%opportunities 
        for further improvements, including noise reduction, 
        %%%ER%%%
        approaches beyond simple PI control
    e.g., 
        adaptivity to workload variations, 
        and fully distributed designs. 
    \end{itemize}

    The rest of the paper is organized as follows. Section~\ref{sec:background} presents the background and related works. Section~\ref{sec:controller_design} details the design and implementation of the controller. Section~\ref{sec:implementation} describes how we evaluated our system and presents the experimental results. Section~\ref{sec:discussion} proposes discussions and identified 
    perspectives %%%ER%%% and improvements 
    before concluding the paper.

    \begin{figure*}
        \centering
        \includegraphics[keepaspectratio,width=0.8\linewidth]{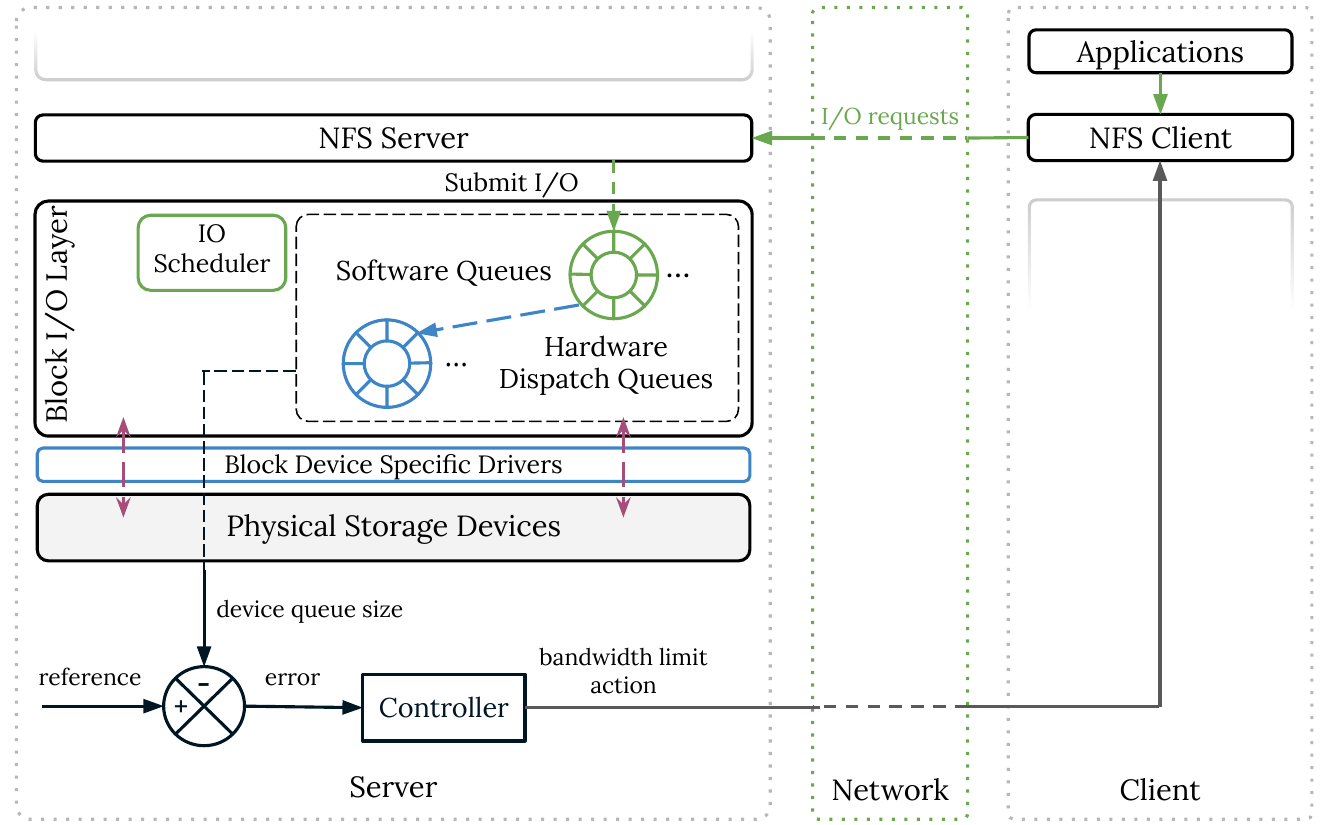}
        \caption{Overview of the control on the I/O path on computing cluster}
        \label{fig:io-stack-control}
    \end{figure*}
    
\section{Background and Motivation}%Motivation / Problem}
\label{sec:background}
    HPC systems are designed to solve large and complex problems by coordinating multiple computing nodes. These computing nodes work collaboratively on tightly coupled tasks, frequently exchanging intermediate data and accessing common datasets. To enable this collaboration, they rely on a shared storage space that must deliver both high throughput and low latency. 
    This shared storage space is usually accessed by the clients via a networked filesystem such as NFS, CephFS, LustreFS, etc.
    
    As an example, Figure~\ref{fig:io-stack-control} depicts an overview of the I/O path in the case of the NFS protocol. NFS allows clients to access files stored on a remote server as if they were local. The NFS client sends file operation requests (open, read, write) over the network to the NFS server, which manages the shared storage space. The server translates these requests into local file system operations. At a lower level, the block I/O layer handles scheduling and queuing of read/write requests, ensuring efficient access to storage. Finally, these requests reach the physical storage devices (e.g., HDDs, SSDs), where data is stored and retrieved, then sent back through the stack to the client.
    The efficiency of this networked filesystem is critical: any bottleneck can slow down the entire computation.

\subsection{I/Os and Congestion}%problem}
    The I/O stack is composed of multiple layers, from the application and file system down to device drivers and storage hardware. Each layer introduces its own policies and optimizations, such as buffering, caching, and scheduling, to improve performance~\cite{io_stack_survey}. However, these mechanisms interact in complex ways, making overall system behavior difficult to predict. When workloads are dynamic and resource contention arises, these interactions can amplify congestion and degrade performance. I/O congestion can stem from multiple sources: the client, the network, the server, or a combination of these~\cite{YildizO2016Root, calciom}. Even more challenging, congestion at one layer can propagate to others, creating contention elsewhere and producing cascading performance issues.

    In this paper, we argue that I/O congestion can be mitigated through dynamic regulation of client-side I/O rates driven by a feedback loop. We advocate that this feedback loop should adopt an end-to-end perspective, managing performance across the entire path from right under the application to the storage device. By abstracting away the complexity of intermediate layers, such an approach enables coordinated adaptation that addresses congestion holistically rather than through layer specific optimizations. 
    
    This approach enables to support a wide range of hardware platforms and system architectures, which fine-tuning every layer of the stack would prevent. In addition, it avoids relying on prior knowledge about I/O profiles or access patterns (see ~\cite{io_patterns_survey} for a comprehensive survey) that might be hard to obtain as these are determined by users at submission. This uncertainty prevents us from pre-configuring system parameters effectively.

\subsection{Feedback Loop}%problem} 
\label{sec:feedback_loop}
    Feedback regulation has been widely used to dynamically monitor computing systems' performance and adjust their behavior to meet desired objectives \cite{Feedback_Control_of_Computing_Systems}. Examples include the MAPE-K loop in autonomic computing \cite{rutten:hal-01285014} and control-theory-based feedback mechanisms \cite{guilloteau:hal-04666859, Software_Engineering_Meets_Control_Theory}.
    
    In our system, a closed-loop control mechanism is employed as shown in the bottom of Figure \ref{fig:io-stack-control}, where we continuously monitor performance at run-time through appropriate sensors (metrics), detailed in Section \ref{sec:control_sensors}. These measurements are compared against a predefined reference state, that generates an error signal, quantifying deviation from  desired  behavior.
    
\begin{figure}
        \centering
        \includegraphics[width=0.85\linewidth]{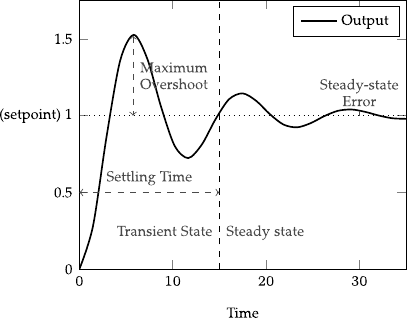}
        \caption{Graphical explanation of the different properties of a controlled system (stability, settling time, overshoot) \cite{article}}
        \label{fig:design_specifications}
    \end{figure}

   To correct the deviation, the controller uses an actuator to dynamically adjust system settings in response to the error signal. The actuator executes the control action, allowing the system to adapt at run-time. This feedback loop operates continuously to achieve the control objectives specified in Figure \ref{fig:design_specifications}, ensuring that the system meets user-defined performance criteria: (i) \emph{reference tracking}, which ensures that the system's output, represented by the black curve in the figure, closely follows the desired reference value with minimal \emph{steady-state error} -- the residual discrepancy at \( t \to \infty \) between  reference and  actual output, (ii) the system must achieve a fast and stable response, meaning the output should reach the reference value with a \emph{short settling time}, defined as the time required for the output to remain within 5\% of its final value, with minimal oscillations and \emph{overshoot}, meaning the system avoids rising excessively above the reference when the input changes.
    
    By achieving these objectives, the system can autonomously optimize performance, maintain stability, and respond efficiently to workload variations and unexpected disturbances. The translation of those objectives in the I/O congestion regulation problem that we consider is explained in the next section, where we present the necessary steps in building such a controller.

\section{Controller Design}
\label{sec:controller_design}

    This section describes the design and implementation of a controller that dynamically regulates client-side I/O rates to reach desired storage activity objectives. It details the common methodology for building a controller, adapted to the selected system. We first need to define the goals that we want to achieve, possible metrics that would help us characterize the disk activity of the server (see Section~\ref{sec:control_sensors}. We then have to select the actions that we will apply on the system to achieve the defined goals. Our choice is explained in Section \ref{sec:control_actuators}. We discuss how these choices affect the implementation of the controller in our cluster environment in Section \ref{sec:control_archi}. After making those choices, it is possible to derive the model that the controller will use, defining the relationship between variations of actions and the selected system metrics. This is explained in Section \ref{sec:control_model}, and reported experimentally in Section \ref{sec:open_loop}. Once a model is elected, we can design the controller, as in choosing which type of controller is appropriate in our context and what are the right parameters for it to obtain good results. This is described in Section~\ref{sec:controller_implem}. Once this is done, it is then possible to test the controller on a real context and validate it, as shown in Section \ref{sec:control_validation}.
    
    Following this methodology, we obtain a control stack that we describe in Figure \ref{fig:io-stack-control} and that we will reference throughout this section.
    
    \subsection{Choosing the sensors}
    \label{sec:control_sensors}
    
        Building a controller requires to identify a sensor, to quantify the congestion level of the shared storage. This measure should be done on the server side where the storage device resides.
        Control Theory requires the chosen sensor to fit multiple criteria for the controller design, e.g. continuous values, precision, etc. Dealing with I/Os also adds some time granularity or delay requirements as the implied I/O mechanisms are fast. The sensor should reflect as closely as possible the state of the storage device at a given time. Looking at the complex I/O stack, we focus on the layers closest to the disk as they are the first to be affected by disk congestion. We opt for not using the disk utilization rate as a congestion indicator because a rate of 100\% does not necessarily mean that the disk is congested, simply that it is treating requests at its maximum speed, which is actually good for performance. Instead, we use the size of the dispatch queue of the targeted block device, on the block layer, to measure disk utilization and congestion level [2]. This queue fills up when I/O requests are waiting for the targeted disk to treat them, indicating that too many requests are sent over to the server. This sensor is illustrated on the left side of Figure \ref{fig:io-stack-control} and is an input of the feedback loop.
    
    \subsection{Choosing the actuators}
    \label{sec:control_actuators}
        We then select an actuator, responsible of dynamically adjusting parameters on the clients to throttle executing workloads.
        
        As well as for the sensor, there are requirements for the actuators that we should try to maintain. Mainly, the actuator needs to have an impact on the system and the measurement that we chose to qualify its state (i.e. dispatch queue size). It also preferably needs to have a short and stable response time and to support a continuous range of command values. On this cluster environment, we have to consider the network layers in the I/O stack. We make the choice to place the actuator on the client side, so that we prevent the network links from getting saturated, and act early on the I/O path to have a better impact on the congestion that we want to avoid. We choose to dynamically adjust an outgoing bandwidth limit on the client nodes as the action. We set it with the Linux tool \texttt{tc}, which allows to define a bandwidth limit using the \textit{Token Bucket Filter}\footnote{\url{https://linux.die.net/man/8/tc-tbf}} algorithm that happens on lower layers of the kernel to manipulate the network packet queues. This actuator can be seen on client side in Figure \ref{fig:io-stack-control}. It does the transition from our controller to the system, by setting the bandwidth limit of the client.
        
    \subsection{Control architecture}%System description}
    \label{sec:control_archi}
        The sensor and controller being on the server node, and the actuators being on the client nodes, we need to establish a one-way communication from the server to the clients in order to share the desired bandwidth limit action. For that, the controller sends the actions in a multicast group to which daemons running on the clients register. These daemons are responsible for updating the bandwidth limit on the host they are when they receive an action from the server. %More advanced ways of transmitting the action to clients are discussed in Section \ref{sec:perspectives}. 
        
        For this approach, and because we study a specific workload that is run across all the clients, we apply the same action on all the clients. We are planning to study the relevance of this choice and how our controller setup can be adapted to support multiple concurrent workloads. We start a reflection about this in Section \ref{sec:portability}.
        
        % Figure \ref{fig:io-stack-control} shows  how the selected sensors and actuators interact with the I/O stack. We also represent the network between the clients and the server to highlight the necessary communications for the actions sent by the server and for the I/O requests emitted by the client applications.
    
    \subsection{Deriving the model}
    \label{sec:control_model}
    In control system design, an open-loop model describes how the system evolves in response to its inputs without any feedback correction. Studying the open loop is important because it shows how the system naturally behaves, whether it is stable, and which parts of its behavior matter most for control. 
    The chosen model should accurately capture the essential dynamics while remaining simple enough for effective control design. To use linear control theory, we approximate the system to a first-order linear model in discrete time, given by:
\begin{equation}
\label{eq:model}
    q(k+1) = a \cdot q(k) + b \cdot \mathrm{bw}(k)
\end{equation}
where $q(k)$ is the dispatch queue size at time $k$, $\mathrm{bw}(k)$ is the bandwidth value at time $k$, and $a, b$ are identified system parameters.

    \subsection{Controller Implementation}
    \label{sec:controller_implem}
        We can now select what type of controller we want to implement. This choice is driven by the properties that we want to achieve (stability, responsiveness, etc.). Some controllers are less complex than others, considering or not past system states, or predicting future ones. Some require a model, or are only data-driven. They are also more or less 
        %%%ER%%%sensible
        sensitive
        to system perturbations such as network fluctuations, or other tasks targeting the shared storage that would affect our measurements while not being considered in our model.  In this work, we do not consider perturbations.

        For the control strategy, we use a discrete-time Proportional-Integral (PI) controller \cite{astrom1995pid}, chosen for its balance between responsiveness and stability. Its simple yet effective design makes it a widely adopted solution, with most industrial systems relying on it. The PI controller computes the next action to be applied on the system based on the error between the measured value of the sensor and its reference value $e[k]$. The proportional term provides immediate correction based on deviations (current error signal $e[k]$), ensuring fast response (reducing settling time). The integral term accumulates past errors to eliminate steady-state error (performance drift from target) over time as illustrated in the following equation:

        \begin{equation}
    \mathrm{bw}(k) = K_p e(k) + K_i T_s \sum_{j=0}^{k} e(j)
        \end{equation}

        Where $K_p \ \text{and} \ K_i$ are the PI controller gains, $e(k)$ is the error at discrete time step k, where k is an integer index corresponding to multiples of the sampling time $T_s$.
        
        The sampling time significantly affects both system stability and performance. If the sampling time is too short, the system becomes unstable due to large signal variations and sensitivity to random noise, while a larger one would cause measurement imprecision due to average on long periods.  Setting the sampling time to $T_s=300ms$ has experimentally shown a good trade-off.
                
        Tuning of the controller gains \(K_p\) and \(K_i\) directly influences the performance objectives -- reference tracking, stability, overshoot, and settling time explained in \cref{sec:feedback_loop}. Proper tuning also improves robustness, enabling the system to handle disturbances and varying workloads more effectively.
        
        To tune the PI controller, we adopt a classic model-based methodology \cite{Borase2020} that, given the approximated linear dynamic model of the system, maps the design specifications into the proportional and integral gains $K_p$ and $K_i$: 
        
            \begin{equation} \label{eq:tuning} 
            \begin{cases} K_p &= \frac{a - r^2}{b} \\ K_i &= \frac{1 - 2r \cos \theta + r^2}{b} 
            \end{cases} 
            \end{equation} 
            where: 
            \begin{equation} 
            \begin{split} r &= \exp\left(-\frac{4T_s}{K_s}\right) \\ \theta &= \pi \, \frac{\log r}{\log M_p} 
            \end{split} 
            \end{equation}
        Here, \(a\) and \(b\) are the system parameters from \cref{eq:model} and \(T_s\) is the sampling time, while \(r \in (0,1)\) and \(\theta \in (0,\pi)\) correspond to the target transient response determined by the settling time \(K_s\) and the overshoot \(M_p\).

    \subsection{Validating the controller}
        \label{sec:control_validation}
        Once every previous step is done, it is possible to finally implement the controller and validate it on the real system. Details on how we implement the sensor, actuator and controller are presented in Section \ref{sec:setup}. Validation is done by applying a wide range of control targets and see how the system reacts to them in terms of convergence, response time, overshoot, etc.

\section{Control Experiments and Results}%Control implementation}
\label{sec:implementation}

The objective of this section is to determine the benefits of using a control theory approach for mitigating congestion issues related to shared storage accesses in the context of a computing workload. Throughout the experiment, a synthetic workload was employed to reproduce the behavior of a write-intensive workload, such as checkpointing, which is widely used in HPC. Checkpointing consists in freezing computations in order to save the state of a task on a physical drive, allowing to restart the task from this state if needed. This type of workload is critical to the global performance of tasks, but it also has an overhead on their execution because of the pauses that are required to do the backup. On top of this, it involves all the computing nodes of a task and is well known for causing congestion on storage because of the large quantity of data that needs to be written. 

The objective of the conducted experiments is to determine whether it is possible to improve performances of write-intensive workloads (as a first approach) by mitigating the congestion of the storage server using Control Theory. We present in this section the implementation of a controller on a real testbed by following the methodology that was explained in the previous section.

The source code for reproducing the following experiments can be found on this Software Heritage archive \footnote{\href{https://archive.softwareheritage.org/swh:1:dir:0ba46bc22e2f1bed6a049c274ce401453f8c9d51;origin=https://gitlab.inria.fr/tcollign/congestion-control.git;visit=swh:1:snp:ac30514101ab9c1befdbc65af87b87d127a4caf0;anchor=swh:1:rev:c216a6b6d78e7bb99b355f83cb6c50b4464115af}{swh:1:dir:0ba46bc22e2f1bed6a049c274ce401453f8c9d51}}.

We firstly define the experimental setup that we use for the experiments in Section \ref{sec:setup}. Then, we present the results of the open loop experiment for the system identification in Section \ref{sec:open_loop}. We present the experimental validation of our controller in Section \ref{sec:validation} and the importance of choosing the right controller gains in Section~\ref{sec:tweak}. We then study its impact on workload performances such as run time and stability (Section \ref{sec:control_perf}) and tail latency (Section \ref{sec:tail_lat}) to conclude on the interest of using control theory in congestion regulation.

    \subsection{Experimental setup}
    \label{sec:setup}
        \begin{table*}[]
        \caption{Configuration of the ecotype cluster of Grid'5000}
        \label{tab:ecotype_conf}
        \begin{tabular}{|c|clll|c|c|c|}
        \hline
        \multirow{2}{*}{Nodes} & \multicolumn{4}{c|}{CPU} & \multirow{2}{*}{Memory} & \multirow{2}{*}{Storage} & \multirow{2}{*}{Network} \\ \cline{2-5}
                                 & \multicolumn{1}{c|}{\#} & \multicolumn{1}{c|}{Name} & \multicolumn{1}{c|}{Cores} & \multicolumn{1}{c|}{Architecture} & & & \\ \hline
        \multicolumn{1}{|l|}{48} & \multicolumn{1}{l|}{2} & \multicolumn{1}{l|}{Intel Xeon E5-2630L v4} & \multicolumn{1}{l|}{10 cores/CPU} & x\_86\_64 & \multicolumn{1}{l|}{128 GiB} & \multicolumn{1}{l|}{400 GB SSD} & \multicolumn{1}{l|}{10 Gbps (SR‑IOV)} \\ \hline
        \end{tabular}
        \end{table*}

        The Grid'5000\footnote{\url{https://www.grid5000.fr}} testbed is used to deploy and test the controller. For each experiment, we reserve $n=16$ computing nodes and one server from the same experimental cluster. The results were obtained on the {\em ecotype} cluster, in which every host has the same configuration, described in Table \ref{tab:ecotype_conf}. A NFSv4 partition is shared across the $n$ clients so that the workloads executed on the clients are targeting the storage server.
        
        For the write-intensive workload, we define a FIO job with sequential write I/Os using large block size and file size that is going to be executed simultaneously on all the computing nodes. The precise FIO settings are described in Listing \ref{fio_workload}. This workload entirely fills up the dispatch queue when no bandwidth restriction is applied on the clients. We say that the queue (or the system) is saturated.

        We use Python for the entire control system by defining multiple main classes for root components of the control (ControllerPI, Sensor, Actuator). They are used to define sub classes, specific to a control configuration, depending on the chosen metrics. This makes the deployment of new controllers easier. We develop a sensor of the length of the dispatch queue using the \texttt{time\_in\_queue} metric from the \textit{sysfs} stat file (\texttt{/sys/block/<dev>/stats}). For the actuator we need a sender and a receiver as the actions are sent through multicast. The receiver side consists in waiting for new messages in the group and in replacing the previous TBF limit with the new bandwidth value.
        
        The controller computes the control parameters $K_{p}$ and $K_{i}$ from the previously established model of the system and the desired properties of the controller, including settling time and overshoot (i.e. the maximum peak value of the system response measured from the objective). It then starts the main loop, consisting of polling sensors and using actuators with the computed action.
        
    \begin{lstlisting}[caption={FIO job configuration},label=fio_workload]
    
    rw=write
    size=4g
    bs=1024k
    numjobs=4
    ioengine=libaio
    iodepth=16
    \end{lstlisting}
    
    \subsection{System analysis}
    \label{sec:open_loop}
    
        \begin{figure}
        \centering
        \begin{subfigure}[b]{1\linewidth}
            \centering
            \includegraphics[width=\linewidth]{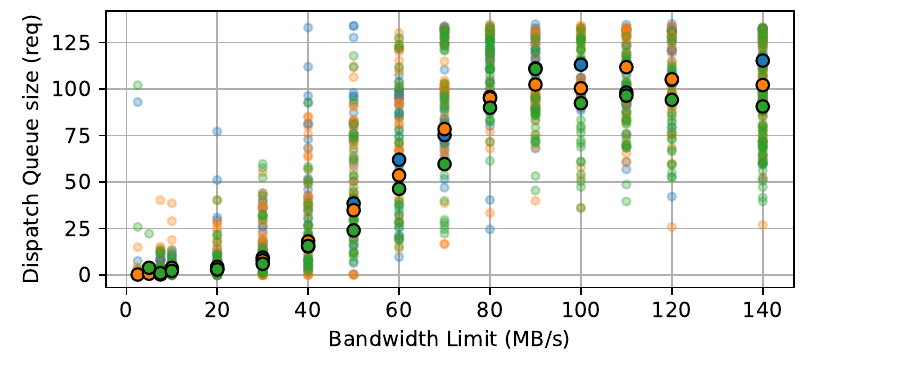}
            \caption{Identification experiments to model Static behavior of the dispatch queue size (y-axis) for fixed values of bandwidth limit input (x-axis). Average dispatch queue size for each bandwidth limit action is displayed for each of the 3 runs (one color per run) for better visualization.}

            \label{fig:openloo_static}
        \end{subfigure}
        \hfill
        \begin{subfigure}[b]{1\linewidth}
            \centering
            \includegraphics[width=\linewidth]{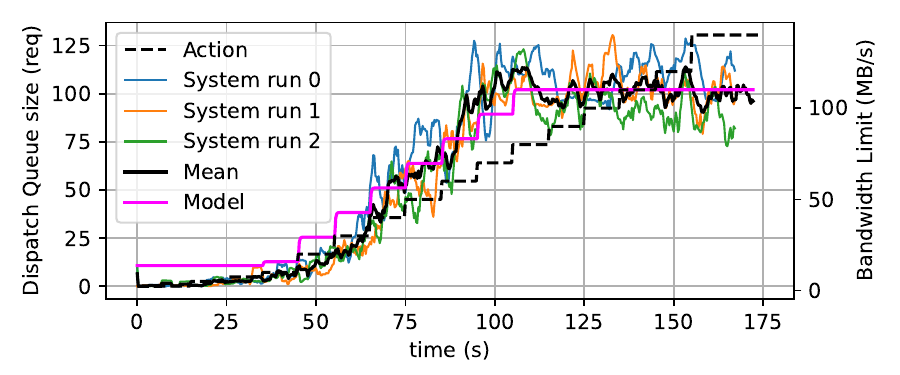}
            \caption{Identification experiments to model dynamic behavior of the dispatch queue size (left y-axis): The model (magenta line) is fitted to the average (black line) of three dispatch queue response to bandwidth limit input variations (dashed black line). We filter the noise of the system with a rolling average over 10 points on the real data and use the effective range of actions to fit the model. The model fits well for certain operation region but relatively poorly for others}
            \label{fig:openloop_dynamic}
        \end{subfigure}
        \caption{Open-Loop Experiments used for Identification -- No Feedback or Control: The upper plot illustrates the system's static behavior under fixed values of input, while the lower plot depicts its dynamic response to input varying during run-time.}
        \label{fig:openloop}
        \end{figure}
        
        The first step before building the type of controller that we chose (PI controller) is to build a model of the studied system, that is, to find a relationship between action and system variations. The system identification results are shown in Figure \ref{fig:openloop}. To understand the structure of the real system we observe the data pattern in the static behaviour (Fig.~\ref{fig:openloop}a) of the dispatch queue size for fixed values of bandwidth limit input.

        We apply a wide range of actions (client bandwidth limits) following an increasing step function and measure the system metric (dispatch queue size) response while executing the selected workload on every client, over multiple iteration. We then fit the first-order linear model described in Section~\ref{sec:controller_design} after filtering the noise with a Savitzky-Golay filter over the collected measures. Also, the data where the queue is saturated and empty are excluded from the fitting phase so that the obtained model better captures the behavior of the queue. On Figure \ref{fig:openloop_dynamic}, we obtain a model (magenta line) that is good-enough to build a controller, it overall follows the behavior of the real size of the dispatch queue while having a decent response time to the action.
    
    \subsection{Control results}
    \label{sec:validation}
        
        This section presents the results obtained with the controller. The goal is to evaluate how effectively our control system can manage to reach any desired system state (level of dispatch queue) in a tolerable amount of time and overshoot.
        
\begin{figure} 
            \centering
            \includegraphics[width=\linewidth]{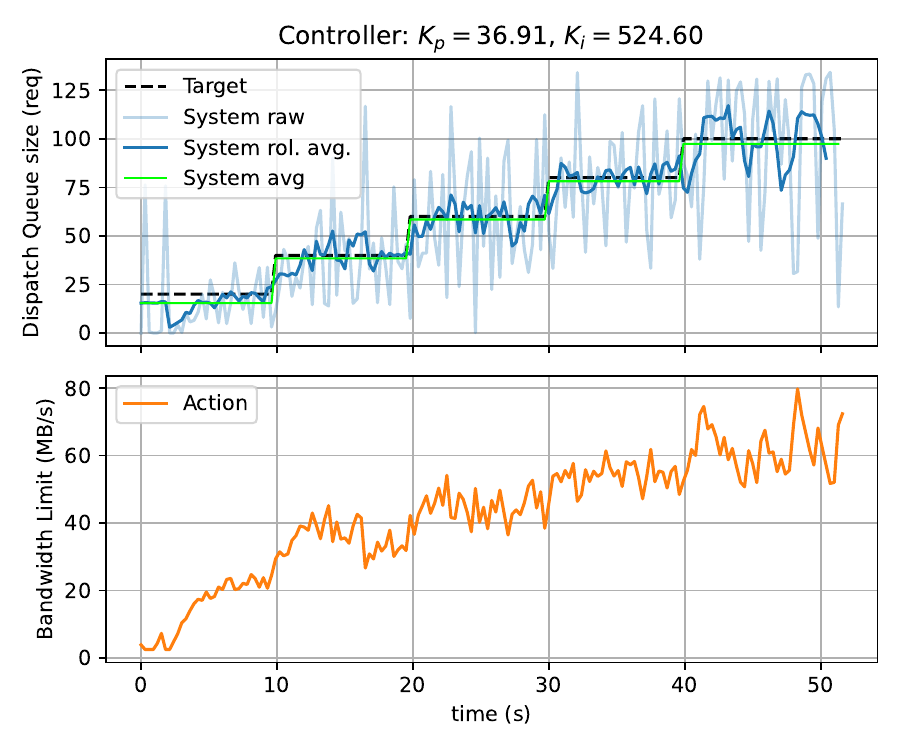}
            \caption{Control results - Top plot represents the dispatch queue size (raw and rolling average) response to control target changes over time. Green line is the average of the system over fixed control targets, showing that the controller manages to reach the desired target on average. Bottom plot is the bandwidth limit action decided by the controller over time}
            \label{fig:control}
        \end{figure}

        For this experiment, we change the control target over time according to a step function. We measure the dispatch queue size (system metric) over time during the execution of the workload and compare it to the desired target. Results are shown in Figure \ref{fig:control}, where the top plot shows the dispatch queue size measured by the sensor as well as a filtered signal with a rolling average, and the control target that that has to be reached. We also plot the average of the system for each segment of fixed targets in order to better visualize the steady state error and accuracy of the control. The bottom plot shows the action computed by the controller over time. We can see that on average, the system follows the desired target despite the noisy behavior. Thanks to the averaged system values, we observe that the steady state error to the target is negligible. The response time is also very short but the noise makes it complicated to determine its value.
        
        About the presence of noise in the results, we can look at the open loop results (Figure \ref{fig:openloop_dynamic}), on which the same rolling averages were applied, and say that the system metric is inherently noisy, so it is to be expected that we observe similar noise during control. We will be discussing the noise mitigation in Section \ref{sec:noise_mitigation}. 
        
        From these results, we see that it is possible to make the system reach any specific state within the tolerated range of targets (limited by the size of the dispatch queue), in a short time. Despite some imprecision of the model and the measured noise, the controller is effective in the context of I/O request regulations, which have a variable and unpredictable behavior.
        
    \subsection{Tuning parameters despite noise}
    \label{sec:tweak}
    
\begin{figure}
        \centering
        \begin{subfigure}[b]{0.49\linewidth}
            \centering
            \includegraphics[width=\textwidth]{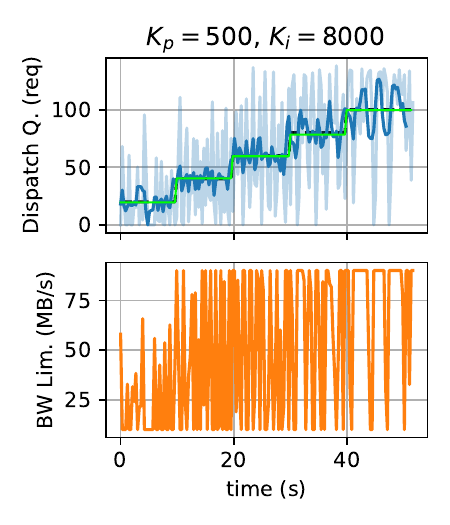}
            \caption{Noisy control results}
            \label{fig:control_noise}
            \vspace{1em}
        \end{subfigure}
        \hfill
        \begin{subfigure}[b]{0.49\linewidth}
            \centering
            \includegraphics[width=\textwidth]{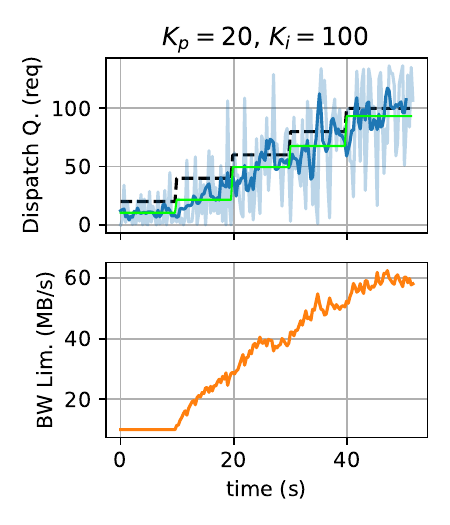}
            \caption{Slow control response and poor reference tracking}
            \label{fig:control_slow}
        \end{subfigure}
        \caption{Control results with multiple control gain configurations. This shows the impact of the controller gains in the quality of the control.}
        \label{fig:control_tweak}
        \end{figure}

   In this subsection, we study the importance of the controller gains ($K_{P}$ and $K_{I}$) in determining control quality, with respect to response time, oscillations, noise, and steady-state error. Figure \ref{fig:control_tweak} shows the results of multiple controller configurations. We observe that the action as well as the system response can be very noisy, as in Subfigure \ref{fig:control_noise}, while still reaching the desired targets on average. Conversely, the action and response can be much less noisy but poorly accurate with respect to the target, with a very slow response, as shown in Subfigure \ref{fig:control_slow} where the control gains are very small. This highlights the importance of finding optimal control gains to mitigate noise while maintaining system responsiveness.

    We argue that the tuning methodology described in \cref{sec:controller_implem} could yield different behavior for two reasons: first, the real system is more complex and it approximated as a linear system around an operating point; second, the system is noisy in open loop, and this noise makes it difficult to be robust and achieve the desired closed-loop specifications. The results shown in Figure \ref{fig:control} were obtained with closed-loop specifications of $M_p = 0.02$ and $K_s = 1.4,\text{s}$.

    \subsection{Performance benefits}
        \label{sec:control_perf}
        
        In this section, we investigate the usefulness of a controller on mitigating the performance degradation caused by congestion. To do this, we run the same FIO workload as previously mentioned without any control and collect the \texttt{job\_runtime} metric of FIO on all the clients. The \texttt{job\_runtime} metric given in the output of FIO only considers the effective part of the workload, excluding the initialization (file creation, etc.) and cleanup phases (flushing I/O requests, etc.). We do the same measurements with a controller for which we define a fixed dispatch queue size target, and we test different values of that target. We repeat each run 5 times to remove any variability and build interpretation of the results.
        
\begin{figure}
            \centering
            \includegraphics[width=\linewidth]{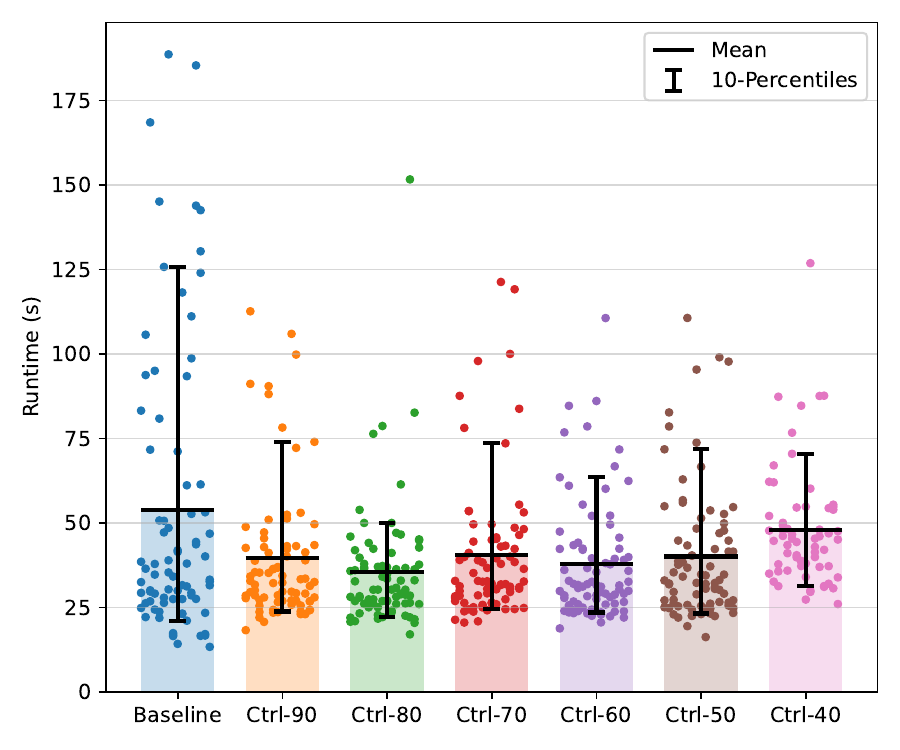}
            \caption{Control performance overhead in comparison of a normal execution of the same FIO workload. Each color represents the execution of the FIO job with a controller whose target is set to a constant dispatch queue size ("Ctrl-90" is for the control with a 90 requests target).}
            \label{fig:control-perf}
        \end{figure}

        Figure \ref{fig:control-perf} shows the results of the execution without control (Baseline) in comparison to controlled executions with different configurations. We plot the job run time for all the clients and runs for the baseline and every control configuration. The mean of all the run times as well as the 10 and 90 percentiles are also displayed. Please note that the outliers across all the runs are not part of a single run, meaning that the disparity in the run times is part of the workload. We see that the choice of the target makes the overall run time of the FIO job vary. Under specific target choice (i.e. 70 or 80 requests in the dispatch queue), the average run time of the jobs on all clients is shorter than for the baseline for a best case scenario of a 20\% average run time improvement with a fixed target of 80 requests. This shows that it is possible to improve the performances of a workload by choosing the right target value.
        
        When looking at the disparity of the points, we can pretend that a good control target choice also leads to a reduced range of run time values across concurrent workloads, improving the overall performance of the workload and stability. Choosing the right target leading to performance improvement is still under investigation and will be discussed in Section \ref{sec:portability}.

        \subsection{Tail latency improvements}
        \label{sec:tail_lat}
        
\begin{figure}
            \centering
            \includegraphics[width=\linewidth]{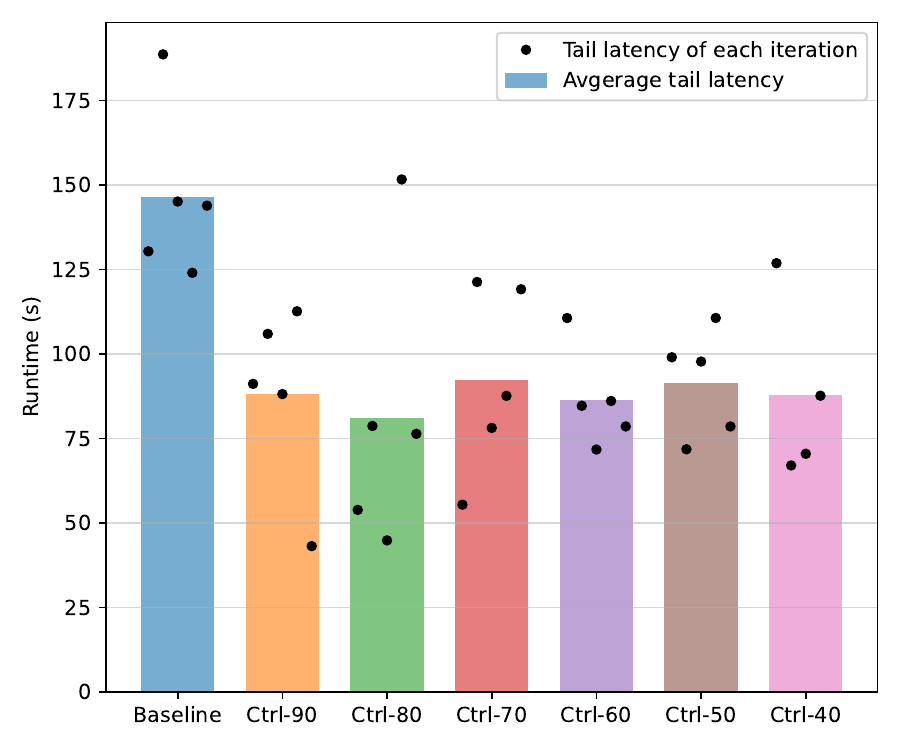}
            \caption{Tail latency of the baseline and multiple control target configuration for each iteration and on average.}
            \label{fig:control-tail}
        \end{figure}

        When considering concurrent processes across multiple hosts, a common performance metric is the tail latency -- the longest run time across all clients -- of the distributed jobs. For a checkpointing process, it corresponds to the pausing time of a task need for a checkpoint. Here, we use the same results as previously mentioned to study the tail latency benefits of our control solution. In Figure \ref{fig:control-tail}, the tail latency of every iteration is displayed as well as an average over all the iterations. For all the tested targets, the tail latency is smaller than the baseline's. A reduction of 35\% of the tail latency of the baseline is achieved with a fixed control target of 70. It is possible to note that reducing the dispatch queue target, which implies to reduce the bandwidth limit of clients, can lead to less variability in the tail latency. This also illustrates the randomness of the consequences of storage congestion on workload performances.

    \section{Discussions and Identified Perspectives}
    
    In this section, we discuss some choices and observations that we encountered throughout the experiments, as well as clearly identified perspectives on how to improve our work and results. Section \ref{sec:noise_mitigation} addresses the topic of noise mitigation. In Section \ref{sec:portability}, we discuss the portability and the workload adaptivity of the control that we implemented and ways to improve them. Finally, in Section \ref{sec:architecure}, we discuss the structural choices that we made in terms of control strategies and hypothesis, and explore possible improvements.
    
    \label{sec:discussion}
    
    \subsection{Noise mitigation}
    \label{sec:noise_mitigation}
    % filter during control
    % understand the noise to add it in the model
    % sampling time
    
    \begin{figure}
        \centering
        \begin{subfigure}[b]{1\linewidth}
            \centering
            \includegraphics[width=\linewidth]{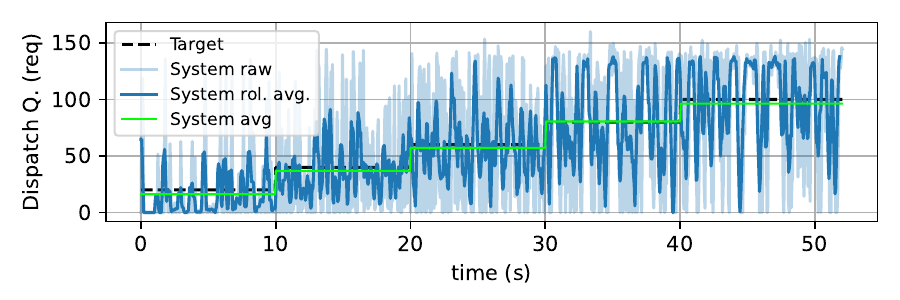}\\[-2ex]
            \caption{$dt=50ms$}
        \end{subfigure}
        \hfill
        \begin{subfigure}[b]{1\linewidth}
            \centering
            \includegraphics[width=\linewidth]{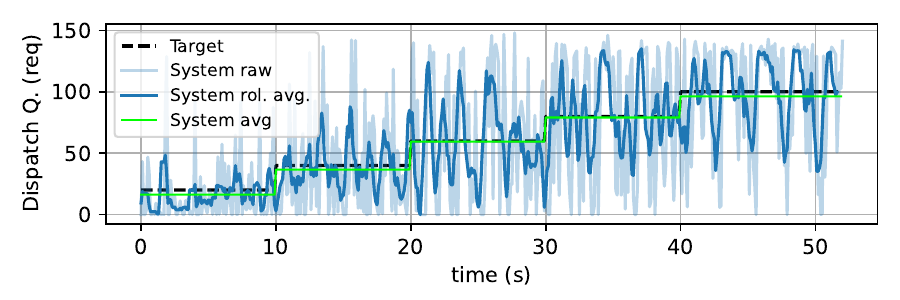}\\[-2ex]
            \caption{$dt=100ms$}
        \end{subfigure}
        \hfill
        \begin{subfigure}[b]{1\linewidth}
            \centering
            \includegraphics[width=\linewidth]{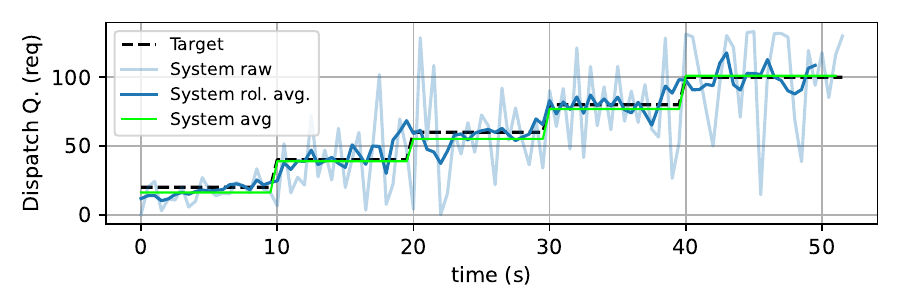}\\[-2ex]
            \caption{$dt=500ms$}
        \end{subfigure}
        \caption{Same controller with different sampling time. The noise amplitude changes with the sampling time.}
        \label{fig:sampling_time}
    \end{figure}

    Throughout the results that we showed in earlier sections, we observe a relatively important noise. Despite showing that this noise is not a response to a poorly defined controller (with similar noise being present during the open loop), possible ways exist to mitigate it. One first approach is to filter the noise out at different levels of the control stack. We implemented an average window on the measurement, or on the measured error. This led to smoother control action depending on the size of the window, but it induces delay in the response as previous measurements are used at each iteration. Furthermore, we sometimes observed oscillations with larger time periods than the noise. In future experiments, we will try new filters such as a Kalman filter, which is widely used in the control theory community.

    We also studied the impact of sampling time on the control result. Figure \ref{fig:sampling_time} compares the control results with the same controller and different sampling times. We see that the noise is effectively smaller as the sampling time is larger, but it comes with a longer response time. This was to be expected as the dispatch queue sensor is in fact based on an average between two consecutive measurements of a base metric (the \texttt{time\_in\_queue} metric). There is a trade-off to find between the accuracy of our measure and its resolution. If the sampling time is too short, the measure will be sensitive to any small variation in our base metric, leading to increased noise and possibly false values. On the opposite, selecting a long sampling time leads to large period averages of the base metric, hiding the smaller high-frequency changes in the system and delaying the system variations.
    
    A possible workaround to avoid having that much noise in the measures but still have a reactive controller would be to have a dynamic sampling time, where the controller would use a short sampling time on target changes and short dispatch queue bursts, and a longer one as the system is stable. This would probably help reduce the noisy behavior on longer phases.
    
    In the end, we see in Figure \ref{fig:sampling_time} that we manage to reach the desired targets on average for any sampling time. Although the noise is often wanted to be attenuated, it might not be necessary in our case. In the context of I/O operations, where there is a lot of variability and unpredictability, with fast-changing I/O loads and bursts, it is likely that filtering the noise out would not be of any interest nor possible. % keep ?

    \subsection{Portability and workload adaptivity}
    \label{sec:portability}
    One of our objectives with this controller solution is to be agnostic to both the architecture and the workload. Our control system is already independent of the architecture it is executing on, the only requirement for deploying the controller on another cluster is to run the open loop experiment and system identification because the change of hardware will likely change the orders of magnitude of the selected metrics and the control parameters. 
    
    About the workload adaptivity of our control system, more reflections need to be made. Until now, we have been doing the open loop and control with a single and identical workload on every client. This workload is also using only one kind of I/O request (write-only requests, fixed block size and I/O depth, etc.), which makes it convenient for building a working controller. We plan to consider more complex workload distributions and variety over clients. Considered strategies include 
    \begin{enumerate*}%[label = (\roman*)]
    \item[(i)] Changing the selected system metric to also consider characterizing properties of I/O requests such as the block size, the type of request (read, write, etc.) or the amount of nodes working for the same job, so that the defined model is more robust and supports more (preferably all) kinds of workload
    \item[(ii)] Using a type of controller that is model-agnostic, meaning that it would only be based on collected data of previous iterations. This would help dealing with the wide variety of workloads.
    \item[(iii)] Moving the controller on the client side for having multiple, separate actions on the clients, this would make the actions more adapted to the system.
    \end{enumerate*}
    On top of this, we want to keep the controller simple to use, with the smallest human intervention possible, which consists up to this date in starting a process that will run an initial open loop phase to divide the model and build the controller, and then start it as a Linux service.
    
    The question of the choice of the optimal control target still remains. It can be found manually, similarly to the study that we did in Section \ref{sec:control_perf}, but that is not a preferable solution since this value would be different for any configuration and workload. Also, having a constant target might not work for workloads with multiple I/O phases, which might require a dynamic target, likely determined by a new performance metric that would be used at run time, such as workload progression, or a more elaborate controller that would activate during intensive I/O phases.

    \subsection{Control architecture choices}
    \label{sec:architecure}
    
    When choosing to have a single workload for every client, we could easily deploy a server-side controller that would decide the allowed outgoing bandwidth of the clients, without differentiating them. This leads to acceptable results in terms of reaching a desired control target, and theoretically maintain fairness across clients in regard to that control. Such action on the system should firstly contain over-consuming clients without degrading the less active ones. When running diverse workloads on clients, it is to be expected that the optimal bandwidths to be allocated to them are going to be different. This rises the question of changing the controller design to support distributed actions. 
    
    As previously mentioned, one possible architecture would be to have a controller on each client, with each controller living independently and making decision for the client it is running on. The used model for the controllers could use server and client side metrics. The server dispatch queue measure could be sent over to clients, in the same way that the action is currently being shared, and additional metrics about I/O requests properties being sent by the NFS client could be measured directly on the clients with tools such as \texttt{eBPF}\footnote{\url{https://ebpf.io}} or \texttt{tcpdump}\footnote{\url{https://www.tcpdump.org/manpages/tcpdump.1.html}}. This would allow to have information on the current state of the storage and on the I/O activity of a client to better determine a suitable bandwidth limit.
    
    In this theoretical setup, we would hope that the independent controllers lead to a stable system behavior that respects control objectives. But it is also possible that the resulting global action of all the controllers will not be appropriate for reaching the control objective, leading to over-throttling and performance degradation, or the opposite. Such control could then benefit from techniques such as agreement protocols or synchronization to allow the system to converge faster towards the desired state.

    \section{Conclusion and Future Works}
    
    In this paper, we addressed the challenges of I/O congestion and performance degradation in distributed clusters by proposing an approach using Control Theory that abstracts the complexity of the I/O stack. We detailed the implementation of a PI controller of the disk activity of a storage server that dynamically adjusts the outgoing bandwidth attributed to computing nodes to reach desired dispatch queue objectives on the server. 
    
    Experimental results carried out on the Grid'5000 testbed validated our approach for a typical write-intensive workload (similar to checkpointing). We reached up to 20\% run time improvement on average and 35\% tail latency reduction for specific fixed control targets, compared to an uncontrolled baseline execution of the same workload.
    
    This study demonstrates that Control Theory is well suited for managing complex I/O dynamics in HPC. % and can be used to mitigate congestion caused by concurrent access to shared storage.
    
    Future works will aim to compare our approach to existing solutions such as Machine Learning or scheduling (e.g. sequentially executing the clients' workloads, removing any degradation caused by concurrence) as well as making the controller handle heterogeneous and more complex workloads. This will require working on more elaborate control strategies such as moving and distributing the controller on the clients, improving the robustness and workload adaptivity of the control model or using model-free approaches.

\begin{acks}
% Pulse
Our work is done in the context of the Inria -- Qarnot Pulse project (see \url{https://www.inria.fr/en/pulse}). It is supported by the ANRT (Association nationale de la recherche et de la technologie) with a CIFRE fellowship granted to Thomas Collignon.
% g5k
Experiments presented in this paper were carried out using the Grid'5000 testbed, supported by a scientific interest group hosted by Inria and including CNRS, RENATER and several universities as well as other organizations (see \url{https://www.grid5000.fr}).
\end{acks}

% \printbibliography
\bibliographystyle{ACM-Reference-Format}
\bibliography{biblio}

%%% -*-BibTeX-*-
%%% Do NOT edit. File created by BibTeX with style
%%% ACM-Reference-Format-Journals [18-Jan-2012].

\begin{thebibliography}{15}

%%% ====================================================================
%%% NOTE TO THE USER: you can override these defaults by providing
%%% customized versions of any of these macros before the \bibliography
%%% command.  Each of them MUST provide its own final punctuation,
%%% except for \shownote{} and \showURL{}.  The latter two
%%% do not use final punctuation, in order to avoid confusing it with
%%% the Web address.
%%%
%%% To suppress output of a particular field, define its macro to expand
%%% to an empty string, or better, \unskip, like this:
%%%
%%% \newcommand{\showURL}[1]{\unskip}   % LaTeX syntax
%%%
%%% \def \showURL #1{\unskip}           % plain TeX syntax
%%%
%%% ====================================================================

\ifx \showCODEN    \undefined \def \showCODEN     #1{\unskip}     \fi
\ifx \showISBNx    \undefined \def \showISBNx     #1{\unskip}     \fi
\ifx \showISBNxiii \undefined \def \showISBNxiii  #1{\unskip}     \fi
\ifx \showISSN     \undefined \def \showISSN      #1{\unskip}     \fi
\ifx \showLCCN     \undefined \def \showLCCN      #1{\unskip}     \fi
\ifx \shownote     \undefined \def \shownote      #1{#1}          \fi
\ifx \showarticletitle \undefined \def \showarticletitle #1{#1}   \fi
\ifx \showURL      \undefined \def \showURL       {\relax}        \fi
% The following commands are used for tagged output and should be
% invisible to TeX
\providecommand\bibfield[2]{#2}
\providecommand\bibinfo[2]{#2}
\providecommand\natexlab[1]{#1}
\providecommand\showeprint[2][]{arXiv:#2}

\bibitem[Astrom(1995)]%
        {astrom1995pid}
\bibfield{author}{\bibinfo{person}{Karl~J Astrom}.}
  \bibinfo{year}{1995}\natexlab{}.
\newblock \showarticletitle{PID controllers: theory, design, and tuning}.
\newblock \bibinfo{journal}{\emph{The international society of measurement and
  control}} (\bibinfo{year}{1995}).
\newblock


\bibitem[Behzad et~al\mbox{.}(2019)]%
        {autotune}
\bibfield{author}{\bibinfo{person}{Babak Behzad}, \bibinfo{person}{Surendra
  Byna}, \bibinfo{person}{Prabhat}, {and} \bibinfo{person}{Marc Snir}.}
  \bibinfo{year}{2019}\natexlab{}.
\newblock \showarticletitle{Optimizing I/O Performance of HPC Applications with
  Autotuning}.
\newblock \bibinfo{journal}{\emph{ACM Trans. Parallel Comput.}}
  \bibinfo{volume}{5}, \bibinfo{number}{4}, Article \bibinfo{articleno}{15}
  (\bibinfo{date}{March} \bibinfo{year}{2019}), \bibinfo{numpages}{27}~pages.
\newblock
\showISSN{2329-4949}
\href{https://doi.org/10.1145/3309205}{doi:\nolinkurl{10.1145/3309205}}


\bibitem[Bez et~al\mbox{.}(2023)]%
        {io_patterns_survey}
\bibfield{author}{\bibinfo{person}{Jean~Luca Bez}, \bibinfo{person}{Suren
  Byna}, {and} \bibinfo{person}{Shadi Ibrahim}.}
  \bibinfo{year}{2023}\natexlab{}.
\newblock \showarticletitle{I/O Access Patterns in HPC Applications: A
  360-Degree Survey}.
\newblock \bibinfo{journal}{\emph{ACM Comput. Surv.}} \bibinfo{volume}{56},
  \bibinfo{number}{2}, Article \bibinfo{articleno}{46} (\bibinfo{date}{Sept.}
  \bibinfo{year}{2023}), \bibinfo{numpages}{41}~pages.
\newblock
\showISSN{0360-0300}
\href{https://doi.org/10.1145/3611007}{doi:\nolinkurl{10.1145/3611007}}


\bibitem[Boito et~al\mbox{.}(2018)]%
        {io_stack_survey}
\bibfield{author}{\bibinfo{person}{Francieli~Zanon Boito},
  \bibinfo{person}{Eduardo~C. Inacio}, \bibinfo{person}{Jean~Luca Bez},
  \bibinfo{person}{Philippe O.~A. Navaux}, \bibinfo{person}{Mario A.~R.
  Dantas}, {and} \bibinfo{person}{Yves Denneulin}.}
  \bibinfo{year}{2018}\natexlab{}.
\newblock \showarticletitle{A Checkpoint of Research on Parallel I/O for
  High-Performance Computing}.
\newblock \bibinfo{journal}{\emph{ACM Comput. Surv.}} \bibinfo{volume}{51},
  \bibinfo{number}{2}, Article \bibinfo{articleno}{23} (\bibinfo{date}{March}
  \bibinfo{year}{2018}), \bibinfo{numpages}{35}~pages.
\newblock
\showISSN{0360-0300}
\href{https://doi.org/10.1145/3152891}{doi:\nolinkurl{10.1145/3152891}}


\bibitem[Borase et~al\mbox{.}(2020)]%
        {Borase2020}
\bibfield{author}{\bibinfo{person}{Rakesh~P. Borase}, \bibinfo{person}{D.~K.
  Maghade}, \bibinfo{person}{S.~Y. Sondkar}, {and} \bibinfo{person}{S.~N.
  Pawar}.} \bibinfo{year}{2020}\natexlab{}.
\newblock \showarticletitle{A review of PID control, tuning methods and
  applications}.
\newblock \bibinfo{journal}{\emph{International Journal of Dynamics and
  Control}} \bibinfo{volume}{9}, \bibinfo{number}{2} (\bibinfo{date}{July}
  \bibinfo{year}{2020}), \bibinfo{pages}{818–827}.
\newblock
\showISSN{2195-2698}
\href{https://doi.org/10.1007/s40435-020-00665-4}{doi:\nolinkurl{10.1007/s40435-020-00665-4}}


\bibitem[Dorier et~al\mbox{.}(2014)]%
        {calciom}
\bibfield{author}{\bibinfo{person}{Matthieu Dorier}, \bibinfo{person}{Gabriel
  Antoniu}, \bibinfo{person}{Rob Ross}, \bibinfo{person}{Dries Kimpe}, {and}
  \bibinfo{person}{Shadi Ibrahim}.} \bibinfo{year}{2014}\natexlab{}.
\newblock \showarticletitle{CALCioM: Mitigating I/O Interference in HPC Systems
  through Cross-Application Coordination}. In \bibinfo{booktitle}{\emph{2014
  IEEE 28th International Parallel and Distributed Processing Symposium}}.
  \bibinfo{pages}{155--164}.
\newblock
\href{https://doi.org/10.1109/IPDPS.2014.27}{doi:\nolinkurl{10.1109/IPDPS.2014.27}}


\bibitem[Filieri et~al\mbox{.}(2015)]%
        {Software_Engineering_Meets_Control_Theory}
\bibfield{author}{\bibinfo{person}{Antonio Filieri}, \bibinfo{person}{Martina
  Maggio}, \bibinfo{person}{Konstantinos Angelopoulos},
  \bibinfo{person}{Nicolas d'Ippolito}, \bibinfo{person}{Ilias
  Gerostathopoulos}, \bibinfo{person}{Andreas~Berndt Hempel},
  \bibinfo{person}{Henry Hoffmann}, \bibinfo{person}{Pooyan Jamshidi},
  \bibinfo{person}{Evangelia Kalyvianaki}, \bibinfo{person}{Cristian Klein},
  {et~al\mbox{.}}} \bibinfo{year}{2015}\natexlab{}.
\newblock \showarticletitle{Software engineering meets control theory}. In
  \bibinfo{booktitle}{\emph{2015 IEEE/ACM 10th International Symposium on
  Software Engineering for Adaptive and Self-Managing Systems}}. IEEE,
  \bibinfo{pages}{71--82}.
\newblock


\bibitem[Gainaru et~al\mbox{.}(2015)]%
        {sched_hpc_congestion}
\bibfield{author}{\bibinfo{person}{Ana Gainaru}, \bibinfo{person}{Guillaume
  Aupy}, \bibinfo{person}{Anne Benoit}, \bibinfo{person}{Franck Cappello},
  \bibinfo{person}{Yves Robert}, {and} \bibinfo{person}{Marc Snir}.}
  \bibinfo{year}{2015}\natexlab{}.
\newblock \showarticletitle{Scheduling the I/O of HPC Applications Under
  Congestion}. In \bibinfo{booktitle}{\emph{2015 IEEE International Parallel
  and Distributed Processing Symposium}}. \bibinfo{pages}{1013--1022}.
\newblock
\href{https://doi.org/10.1109/IPDPS.2015.116}{doi:\nolinkurl{10.1109/IPDPS.2015.116}}


\bibitem[Guilloteau et~al\mbox{.}(2024)]%
        {guilloteau:hal-04666859}
\bibfield{author}{\bibinfo{person}{Quentin Guilloteau}, \bibinfo{person}{Sophie
  Cerf}, \bibinfo{person}{Rapha{\"e}l Bleuse}, \bibinfo{person}{Bogdan Robu},
  {and} \bibinfo{person}{Eric Rutten}.} \bibinfo{year}{2024}\natexlab{}.
\newblock \showarticletitle{{Under Control: A Control Theory Introduction for
  Computer Scientists}}. In \bibinfo{booktitle}{\emph{{ACSOS 2024 - 5th IEEE
  International Conference on Autonomic Computing and Self-Organizing Systems
  (ACSOS 2024)}}}. \bibinfo{publisher}{{IEEE}}, \bibinfo{address}{Aahrus,
  Denmark}, \bibinfo{pages}{1--10}.
\newblock
\urldef\tempurl%
\url{https://hal.science/hal-04666859}
\showURL{%
\tempurl}


\bibitem[Hellerstein et~al\mbox{.}(2004)]%
        {Feedback_Control_of_Computing_Systems}
\bibfield{author}{\bibinfo{person}{Joseph~L. Hellerstein},
  \bibinfo{person}{Yixin Diao}, \bibinfo{person}{Sujay Parekh}, {and}
  \bibinfo{person}{Dawn~M. Tilbury}.} \bibinfo{year}{2004}\natexlab{}.
\newblock \bibinfo{booktitle}{\emph{Feedback Control of Computing Systems}}.
\newblock \bibinfo{publisher}{John Wiley \& Sons, Inc.},
  \bibinfo{address}{Hoboken, NJ, USA}.
\newblock
\showISBNx{047126637X}


\bibitem[Rutten et~al\mbox{.}(2018)]%
        {rutten:hal-01285014}
\bibfield{author}{\bibinfo{person}{Eric Rutten}, \bibinfo{person}{Nicolas
  Marchand}, {and} \bibinfo{person}{Daniel Simon}.}
  \bibinfo{year}{2018}\natexlab{}.
\newblock \showarticletitle{{Feedback Control as MAPE-K loop in Autonomic
  Computing}}.
\newblock In \bibinfo{booktitle}{\emph{{Software Engineering for Self-Adaptive
  Systems III. Assurances.}}} \bibinfo{series}{Lecture Notes in Computer
  Science}, Vol.~\bibinfo{volume}{9640}. \bibinfo{publisher}{{Springer}},
  \bibinfo{pages}{349--373}.
\newblock
\href{https://doi.org/10.1007/978-3-319-74183-3\_12}{doi:\nolinkurl{10.1007/978-3-319-74183-3\_12}}


\bibitem[Shevtsov et~al\mbox{.}(2017)]%
        {article}
\bibfield{author}{\bibinfo{person}{Stepan Shevtsov}, \bibinfo{person}{Mihaly
  Berekmeri}, \bibinfo{person}{Danny Weyns}, {and} \bibinfo{person}{Martina
  Maggio}.} \bibinfo{year}{2017}\natexlab{}.
\newblock \showarticletitle{Control-Theoretical Software Adaptation: A
  Systematic Literature Review}.
\newblock \bibinfo{journal}{\emph{IEEE Transactions on Software Engineering}}
  \bibinfo{volume}{PP} (\bibinfo{date}{05} \bibinfo{year}{2017}),
  \bibinfo{pages}{1--1}.
\newblock
\href{https://doi.org/10.1109/TSE.2017.2704579}{doi:\nolinkurl{10.1109/TSE.2017.2704579}}


\bibitem[Tarasov et~al\mbox{.}(2016)]%
        {filebench}
\bibfield{author}{\bibinfo{person}{Vasily Tarasov}, \bibinfo{person}{Erez
  Zadok}, {and} \bibinfo{person}{Spencer Shepler}.}
  \bibinfo{year}{2016}\natexlab{}.
\newblock \showarticletitle{Filebench: A Flexible Framework for File System
  Benchmarking}.
\newblock \bibinfo{journal}{\emph{login Usenix Mag.}}  \bibinfo{volume}{41}
  (\bibinfo{year}{2016}).
\newblock
\urldef\tempurl%
\url{https://api.semanticscholar.org/CorpusID:56553130}
\showURL{%
\tempurl}


\bibitem[Tipu et~al\mbox{.}(2022)]%
        {autotune2}
\bibfield{author}{\bibinfo{person}{Abdul Jabbar~Saeed Tipu},
  \bibinfo{person}{P\'{a}draig~\'{O} Conbhu\'{\i}}, {and} \bibinfo{person}{Enda
  Howley}.} \bibinfo{year}{2022}\natexlab{}.
\newblock \showarticletitle{Artificial neural networks based predictions
  towards the auto-tuning and optimization of parallel IO bandwidth in HPC
  system}.
\newblock \bibinfo{journal}{\emph{Cluster Computing}} \bibinfo{volume}{27},
  \bibinfo{number}{1} (\bibinfo{date}{Dec.} \bibinfo{year}{2022}),
  \bibinfo{pages}{71–90}.
\newblock
\showISSN{1386-7857}
\href{https://doi.org/10.1007/s10586-022-03814-w}{doi:\nolinkurl{10.1007/s10586-022-03814-w}}


\bibitem[Yildiz et~al\mbox{.}(2016)]%
        {YildizO2016Root}
\bibfield{author}{\bibinfo{person}{Orcun Yildiz}, \bibinfo{person}{Matthieu
  Dorier}, \bibinfo{person}{Shadi Ibrahim}, \bibinfo{person}{Robert~B. Ross},
  {and} \bibinfo{person}{Gabriel Antoniu}.} \bibinfo{year}{2016}\natexlab{}.
\newblock \showarticletitle{{O}n the {R}oot {C}auses of {C}ross-{A}pplication
  {I/O} {I}nterference in {HPC} {S}torage {S}ystems}. In
  \bibinfo{booktitle}{\emph{{IPDPS}}}. \bibinfo{publisher}{{IEEE}},
  \bibinfo{pages}{750--759}.
\newblock
\showISSN{1530-2075}
\href{https://doi.org/10.1109/IPDPS.2016.50}{doi:\nolinkurl{10.1109/IPDPS.2016.50}}


\end{thebibliography}

\end{document}